\documentclass[]{arXiv_v1}

\preprintnumber{XXXX-XXXX} 
\usepackage{hyperref}
\usepackage{comment}
\usepackage{enumitem}

\begin{document}

\title{Modification on thermal motion in Geant4 for neutron capture simulation in Gadolinium loaded water}

\author[1,*]{Y.~Hino} 
\author[2]{K.~Abe}
\author[3]{R.~Asaka}
\author[4]{S.~Han\footnote{now at Department of Physics, Kyoto University, Kyoto 606-8502, Japan}}
\author[2]{M.~Harada}
\author[3]{M.~Ishitsuka}
\author[3]{H.~Ito}
\author[5]{S.~Izumiyama\footnote{now at Kobayashi-Masukawa Institute for the Origin of Particles and the Universe, Nagoya University, Nagoya 464-8602, Japan}}
\author[2]{Y.~Kanemura}
\author[1]{Y.~Koshio}
\author[1]{F.~Nakanishi} 
\author[2]{H.~Sekiya}
\author[2]{T.~Yano}

\affil[1]{Department of Physics, Okayama University, Okayama 700-8530, Japan}
\affil[2]{Kamioka Observatory, Institute for Cosmic Ray Research, University of Tokyo, Kamioka, Gifu 506-1205, Japan}
\affil[3]{Department of Physics, Faculty of Science and Technology, Tokyo University of Science, Chiba 278-8510, Japan}
\affil[4]{Research Center for Cosmic Neutrinos, Institute for Cosmic Ray Research, University of Tokyo, Kashiwa, Chiba 277-8582, Japan}
\affil[5]{Department of Physics, Tokyo Institute of Technology, Tokyo 152-8551, Japan
\email{yotahino@okayama-u.ac.jp}
}



\begin{abstract} 
Neutron tagging is a fundamental technique for electron anti-neutrino detection via the inverse beta decay channel.
A reported discrepancy in neutron detection efficiency between observational data and simulation predictions prompted an investigation into neutron capture modeling in Geant4.
The study revealed that an overestimation of the thermal motion of hydrogen atoms in Geant4 impacts the fraction of captured nuclei.
By manually modifying the Geant4 implementation, the simulation results align with calculations based on evaluated nuclear data and show good agreement with observables derived from the SK-Gd data.
\end{abstract}

\subjectindex{xxxx, xxx}

\maketitle

\section{Introduction}
\label{sec:intro}

Neutron detection by thermal neutron capture on nuclei is a powerful tool for identifying neutrino-induced events in various neutrino experiments.
In the search for anti-electron neutrinos, a delayed coincidence technique of inverse beta decay (IBD), $\bar{\nu}_{e} + p \to e^{+} + n$, is often used, leading to a large suppression of accidental backgrounds without an associated neutron.

Super-Kamiokande (SK), a pure water Cerenkov detector with a fiducial volume of 22.5~kton, performed a search for diffuse supernova neutrino background (DSNB) using IBD by detecting a 2.2~MeV gamma-ray from neutron capture on hydrogen (H), and succeeded in giving a strict upper limit and excluding some optimistic DSNB scenarios~\cite{cite:dsnb_sk4}.
Further background suppression and an increase in the detection efficiency of IBD can be achieved by loading gadolinium (Gd) into water due to the higher cross section and total energy of emitted gamma-rays ($\sim$8~MeV) of Gd capture than those of H capture~\cite{cite:beacom}; thus, SK was upgraded to SK-Gd by dissolving Gd up to 0.01\% (SK-VI)~\cite{cite:first} and 0.03\% (SK-VII)~\cite{cite:second} sequentially.

However, the simulated prediction of neutron detection efficiency has been reported to show a +8.5\% disagreement with calibration data using a neutron source ${}^{241}$americium-beryllium (AmBe)~\cite{cite:harada_ichep2022}.
This fact causes one of the systematic uncertainties not only in signal prediction, but also in background estimation in the recent DSNB search in SK-Gd~\cite{cite:dsnb_sk6}.
The AmBe calibration indicates that the Monte Carlo (MC) simulation with Geant4 simulation toolkit~\cite{cite:geant4_1, cite:geant4_2, cite:geant4_3} underestimates a fraction of H capture in the total neutron capture, i.e., overestimates the Gd capture fraction, compared to the data estimate~\cite{cite:han_thesis}.
This can cause the data-MC discrepancy in the neutron detection efficiency.

Therefore, we have conducted a survey of the neutron capture simulation in the Gd-loaded water based on Geant4 and found a possible modification to improve the understanding of the neutron capture behavior and data reproducibility.
The validation of the modification is discussed based on the observation in the SK-Gd experiment.

\section{Simulation}
\label{sec:sim}
The validation of the neutron behavior in Gd-loaded water in Geant4 has been done with SKG4, which is the Geant4-based SK detector MC simulation~\cite{cite:skg4} and has been used in recent works in the SK-Gd phase (\cite{cite:dsnb_sk6, cite:ncqe_sk6} as examples).
For high-precision neutron transport below 20 MeV, \verb|G4ParticleHP|, the generalized class of \verb|G4NeutronHP|~\cite{cite:G4HP} to unify the treatment of a neutron and the other particles is enabled in SKG4.
Based on evaluated nuclear data libraries, the model simulates neutron behavior with elastic and inelastic scattering, fission, and capture processes in the material.
Neutron thermal scattering is handled by \verb|G4NeutronHPThermalScattering| class when the kinetic energy is less than 4~eV.
In particular, Geant4.10.5.p01, on which SKG4 currently depends, is compatible with \verb|G4NDL4.5| as the neutron data library, which primarily refers to ENDF/B-VII.1~\cite{cite:endf7}.

The detector material in the simulation consists of Gd-loaded water with thermal scattering enabled elements, e.g., \verb|TS_H_of_Water| in $\mathrm{H_{2}O}$.
We implemented an AmBe calibration source in SKG4 for the neutron simulation.
The source consists of an encapsulated AmBe source surrounded by cubic BGO scintillator crystals~\cite{cite:first, cite:harada_thesis}, which allows us to identify a neutron emission of AmBe with a large scintillation light emission of a 4.4~MeV gamma-ray associated with the neutron.
Note that a neutron generated at the center of the SK detector interacts only with the materials of the Gd-loaded water and the AmBe calibration source.
The SK detector is large enough and the other detector components are $\sim$20 m away from the source.

\subsection{Neutron data validation}
First, we performed a validation of the capture cross section data implemented in Geant4 by comparing the value obtained by the \verb|GetXsec| method in the \verb|G4ParticleHPCapture| class with one retrieved from the ENDF/B-VII.1 database as a function of the true neutron velocity.

\begin{figure}[h]
    \centering
    \includegraphics[width=0.7\linewidth]{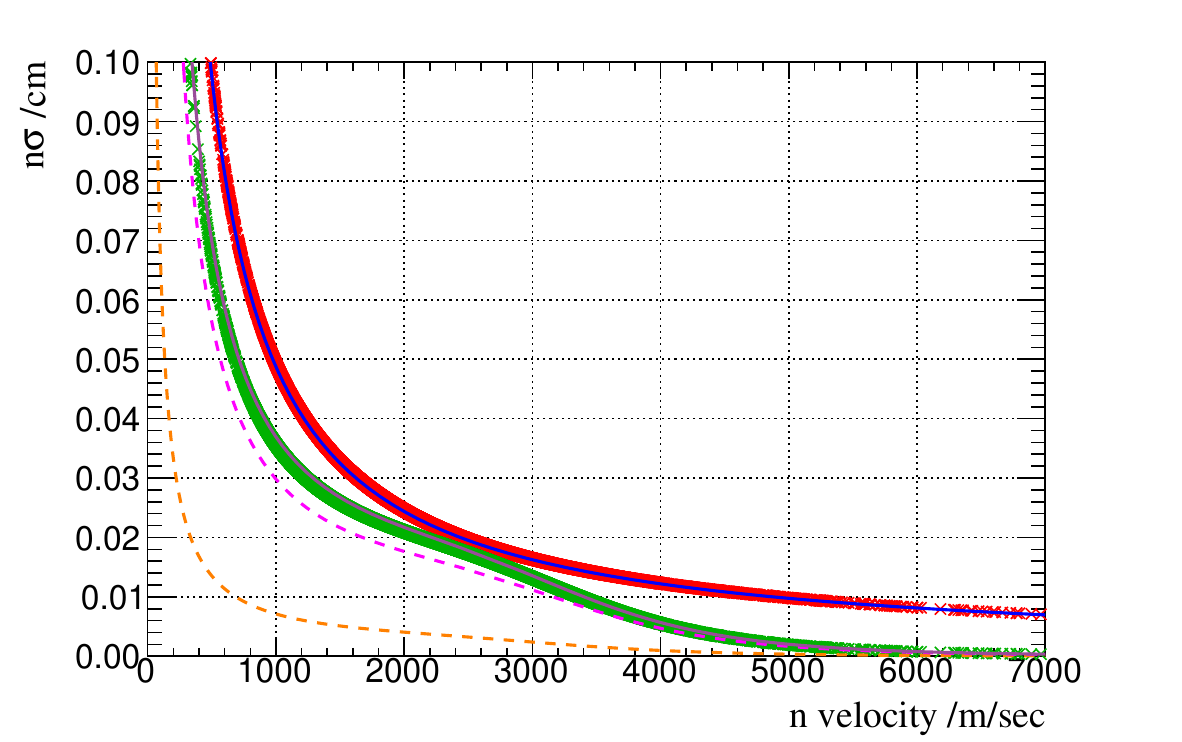}
    \caption{The product of number density and capture cross section $n\sigma$ of H (red) and Gd (green) as a function of neutron velocity dumped from the Geant4. The solid lines correspond to the $n\sigma$ of H (blue) and ${}^{\mathrm{nat}}$Gd capture (magenta) whose cross section values are obtained from the ENDF-VII.1 database. The dashed lines are the $n\sigma$ of ${}^{155}$Gd (orange) and ${}^{157}$Gd (pink), respectively. We assumed 0.0271\% Gd sulfate concentration in these plots.}
    \label{fig:nsig_true}
\end{figure}

Figure~\ref{fig:nsig_true} shows the product of the number density and the capture cross section $n\sigma$ as a function of the true neutron velocity.
The Geant4 output (marker) is identical to the calculation based on the capture cross section from ENDF/B-VII.1 (solid line) for both H and Gd.
In particular, the cross section values at 2200~m/s, a mean of the thermal neutron velocity following the Maxwellian distribution, from each of them are listed in Table~\ref{tab:xs_dump} which shows identical results for each element.

\begin{table}[b]
    \centering
    \caption{Capture cross section of each nucleus dumped from Geant4 (G4NDL4.5) and retrieved from the ENDF-VII.1 data library at the neutron velocity 2200~m/s.}
    \begin{tabular}{cccc}
    \hline \hline
     Source            & $\sigma_{\mathrm{H}}(v_{\mathrm{therm}})$ /b & $\sigma_{\mathrm{{}^{155}Gd}}(v_{\mathrm{therm}})$ /kb & $\sigma_{\mathrm{{}^{157}Gd}}(v_{\mathrm{therm}})$ /kb \\ \hline
     Geant4 (G4NDL4.5) & 0.332 & $60.7$ & $253.0$ \\
     ENDF/B-VII.1      & 0.332 & $60.5$ & $253.6$ \\ \hline
    \end{tabular}
    \label{tab:xs_dump}
\end{table}

Based on this survey, we concluded that the evaluated data was correctly implemented in \verb|G4NDL4.5| and that the cross section was unlikely to cause the discrepancy in neutron detection efficiency between the data and MC.

\subsection{Thermal boost}
Due to the kinetic energy of thermal neutrons, the effect of the thermal motion of the interacting material is essential for thermal neutron transport. 
An effective neutron velocity with respect to the material with thermal motion is computed by the \verb|GetThermalEnergy| method in the \verb|G4ParticleHPThermalBoost| class, which is commonly included in the interaction classes (elastic, inelastic, capture and fission) of \verb|G4ParticleHP|.
In the interaction classes, \verb|GetXsec| is called as a function of the relative kinetic energy instead of the kinetic energy in the laboratory frame.
The thermal motion calculation is based on the free gas approximation of a target nucleus and is performed in the \verb|GetBiasedThermalNucleus| method of the \verb|G4Nucleus| class.
The approximation assumes that neutrons perceive the material as a free gas of nuclei with a Maxwell velocity distribution at a given temperature.
Indeed, this picture would work in a fluid, such as the Gd-loaded water.

\begin{figure}[th]
    \centering
    \includegraphics[width=0.49\linewidth]{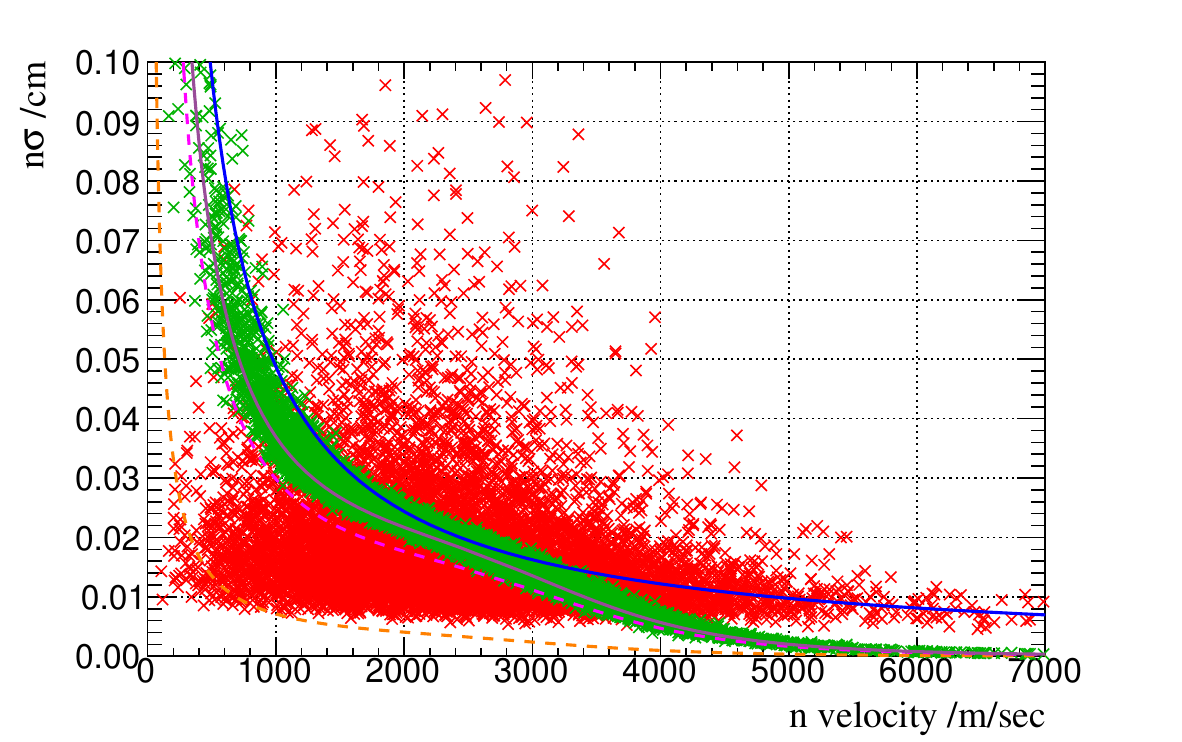}
    \includegraphics[width=0.49\linewidth]{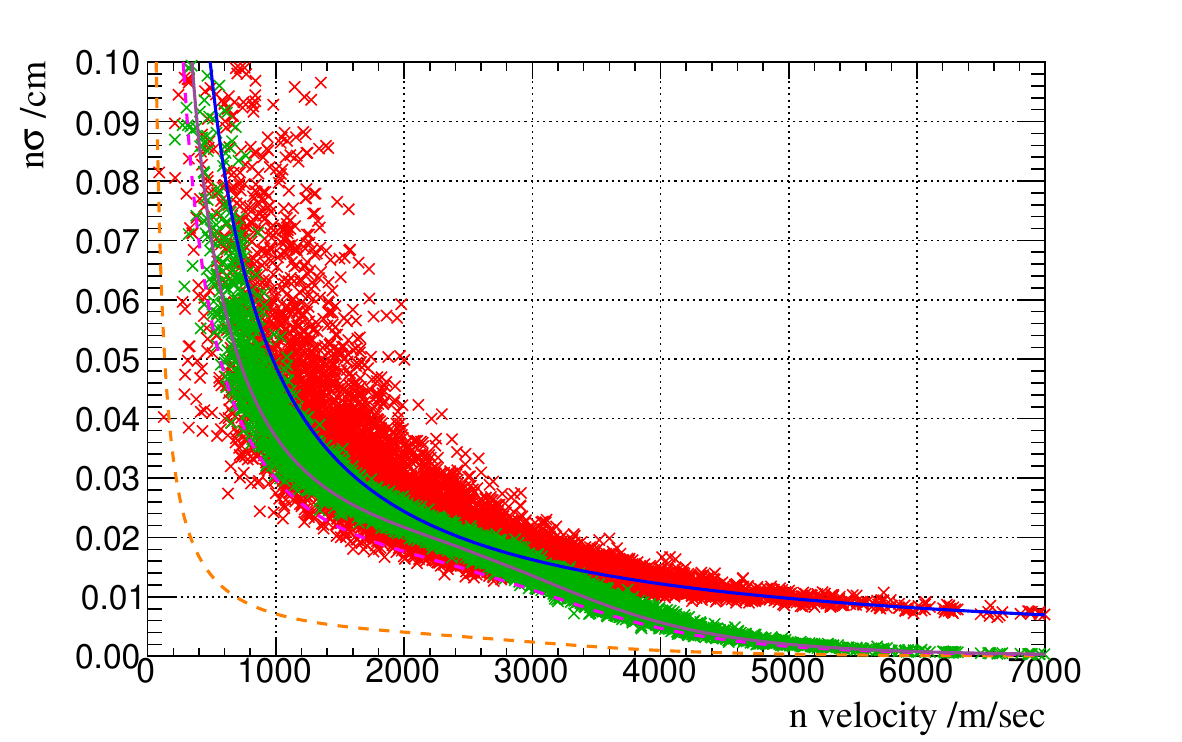}
    \caption{Correlation between $n\sigma(v^{e}_{\mathrm{rel}})$ of H (red) and Gd (green)  as a function of the true neutron velocity dumped from the default Geant4 (left) and the Geant4 with the thermal boost modification in the hydrogen capture case (right). The solid and the dashed lines are identical to the lines in Fig.~\ref{fig:nsig_true}. 0.0271\% Gd sulfate concentration is assumed in these plots.}
    \label{fig:nsig_therm}
\end{figure}

However, we found an unexpected simulation result when the target nucleus is hydrogen.
Figure~\ref{fig:nsig_therm} (left) shows the product of the number density and the capture cross section at the relative neutron velocity with respect to a target nucleus, $n\sigma(v^{e}_{\mathrm{rel}})$ ($e =$ H, Gd), as a function of the true neutron velocity.
Due to the randomization that takes into account the thermal motion of the material, it is natural that $n\sigma(v^{e}_{\mathrm{rel}})$ should be distributed around the curves shown in Fig.~\ref{fig:nsig_true}.
On the other hand, the distribution is spread widely and a large fraction of H capture (red marker) shows a smaller value than Gd capture (green) does in the thermal neutron velocity region.
This strongly suggests why the fraction of H capture is underestimated in Geant4.

We have found that the problem comes from an implementation of material in Geant4. 
Geant4 recognizes a material as a set of elements, and no information about a unit of the material, i.e., a molecule, is considered in the calculation. 
Thus, the thermal motion of the target material computed in \verb|GetBiasedThermalNucleus| considers it as a monoatomic molecule.
This approximation fortunately works if the target material is Gd, because Gd exists as an ion, Gd${}^{3+}$, in water and behaves as a free monoatomic molecule.
In contrast, H is bound in H${}_{2}$O molecular in water, and the mass of H${}_{2}$O should be considered in the thermal motion instead of the mass of the H element.

Due to the difficulty of the generally implementing an algorithm that takes into consideration the state of each element in the material because of chemical complications, we added a conditional branch in \verb|G4ParticleHPThermalBoost.hh| as a specific but simple prescription for water.
The \verb|GetBiasedThermalNucleus| method calculates the thermal motion of a target nucleus as a function of its mass and material temperature.
Thus, we have replaced the mass with the mass of H${}_{2}$O if the target is hydrogen.
Figure~\ref{fig:nsig_therm} (right) shows the result of this prescription in $n\sigma(v^{e}_{r})$ as a function of the true neutron velocity, in a manner analogous to that of the left plot. 
It is obvious that $n\sigma(v^{\mathrm{H}}_{\mathrm{rel}})$ is reasonably distributed along the $n\sigma$ curve from the ENDF-VII.1 database as a result of the modification, unlike the original.
This indicates that the underestimation of the H capture probability is corrected.
The following section shows how the modification of the thermal boost treatment of hydrogen is validated based on observables in SK-Gd.

\section{Result}
\label{sec:result}
In order to validate our modification to the thermal boost treatment of H, observables in SK-Gd, such as the capture time constant and the H capture fraction, are compared with the analytical calculation and the observed data.
The MC outputs of the observables were extracted from MC truth information, i.e., no full detector simulation, including event reconstruction, was performed.

\subsection{Analytical calculation}
Based on the cross section retrieved from the evaluated nuclear data library, the observables can be simply computed.
The capture time constant $\tau$ is written as follows:
\begin{equation}
    \tau = \frac{1}{\sum_i n_{i}g_{w}^{i}(T)\sigma_{i}^{\mathrm{therm}}v_{\mathrm{therm}}},
    \label{eq:capt}
\end{equation}
where $n_i$ is the number density, and $g_{w}^{i}(T)$ is the Wescott g-factor at temperature $T$ of an element $i~(= \mathrm{H, Gd})$. 
The capture cross section is averaged over the Maxwellian distribution with the mean value of the thermal neutron velocity, i.e., $v_{\mathrm{therm}} = 2200$~m/s, as expressed in $\sigma^{\mathrm{therm}}_{i}$.

Accordingly, the H capture fraction $R_{\mathrm{H}}$ can be computed using the same parameters and variables in
\begin{equation}
    R_{\mathrm{H}} = \frac{n_{\mathrm{H}}g_{w}^{\mathrm{H}}(T)\sigma_{\mathrm{H}}^{\mathrm{therm}}}{\sum_i n_{i}g_{w}^{i}(T)\sigma_{i}^{\mathrm{therm}}}.
    \label{eq:hfrac}
\end{equation}
Note that due to the negligible contribution from capture on the other nuclei, the fraction of Gd capture is simply approximated as $R_{\mathrm{Gd}} = 1 - R_{\mathrm{H}}$, since capture on oxygen is negligible.
For the calculation, the Maxwellian averaged cross section and the Wescott g-factor values for each element are taken from the ENDF-VII.1 table and the reference~\cite{cite:gfac}, respectively, and the number densities are estimated for a given Gd concentration in the Gd-loaded water.

\subsection{Capture time constant}
\label{sec:result_capt}
\begin{figure}[h]
    \centering
    \includegraphics[width=0.49\linewidth]{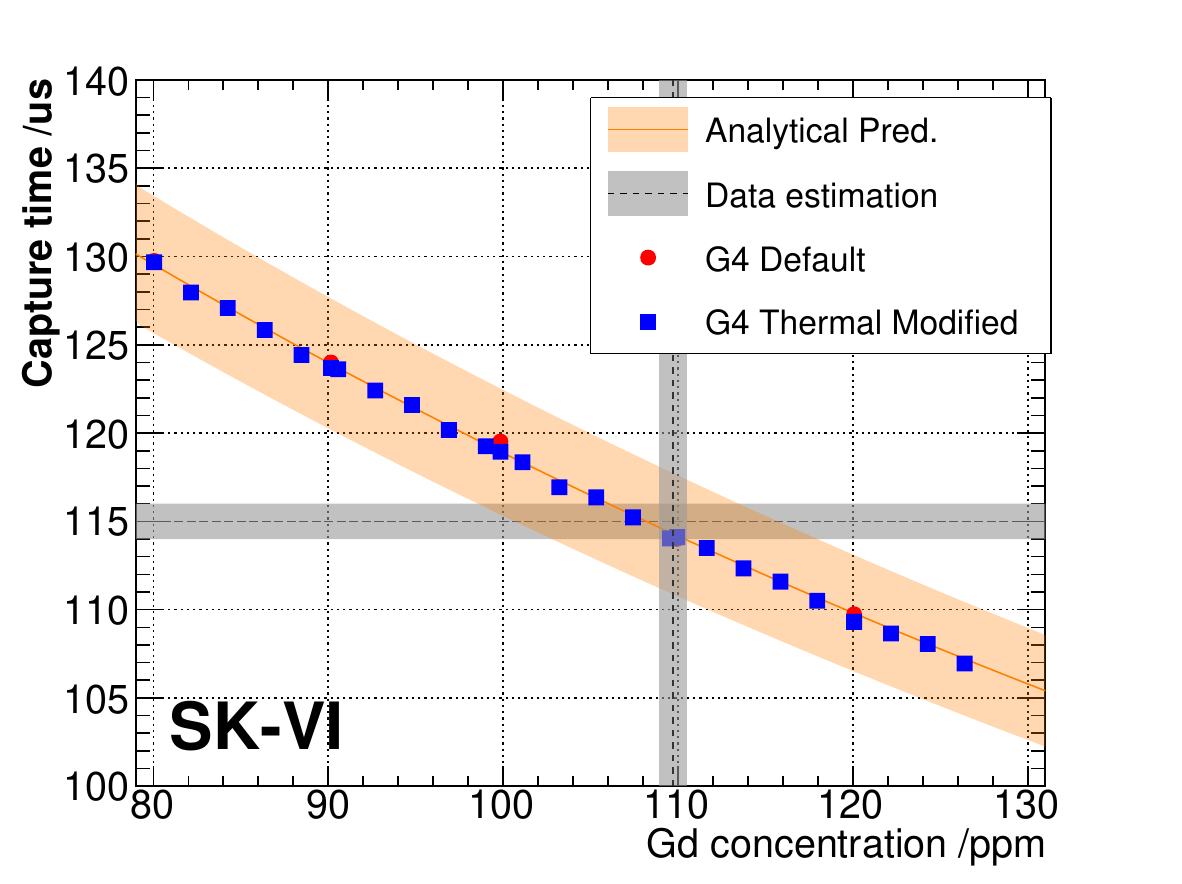}
    \includegraphics[width=0.49\linewidth]{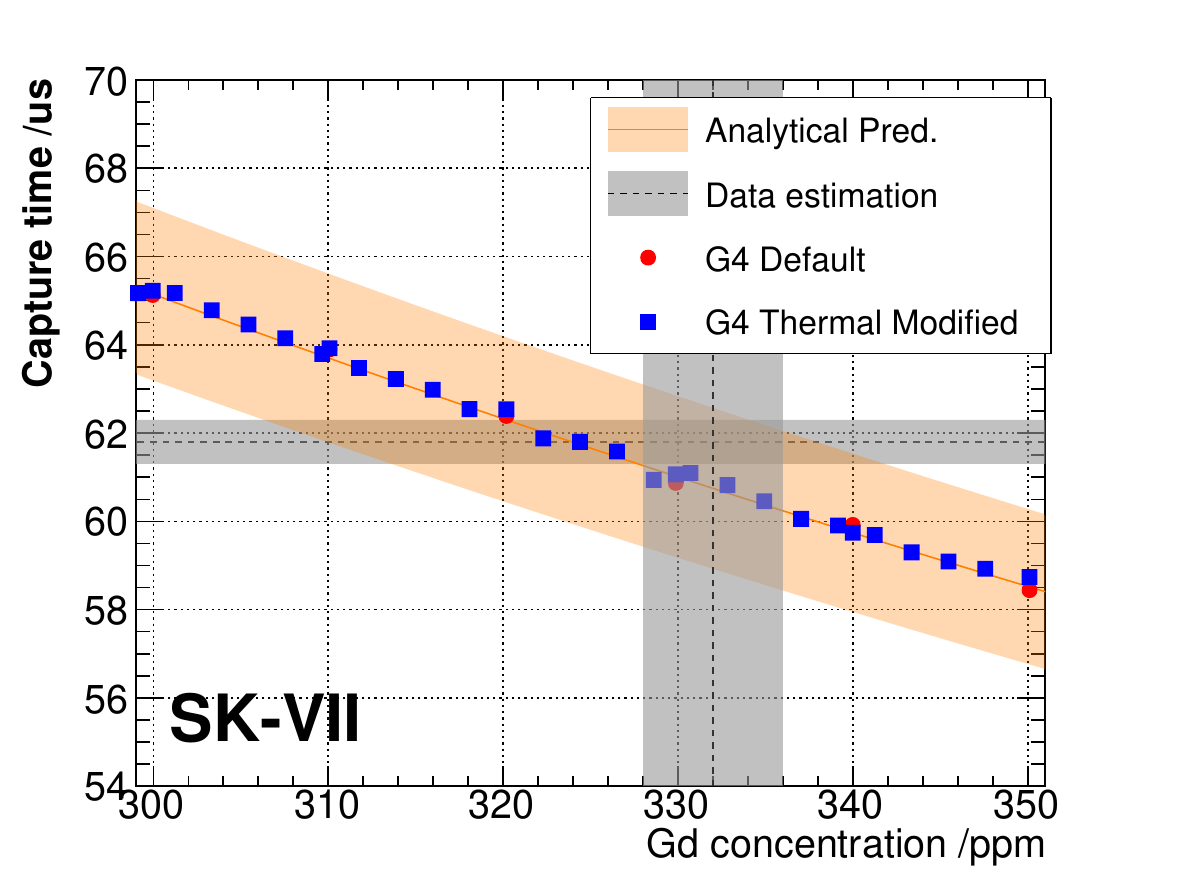}
    \caption{Capture time constant as a function of the Gd concentration around SK-VI 0.01\% (left) and SK-VII 0.03\% (right). Both the default (red circle) and the modified Geant4 (blue square) give identical results to the analytical calculation (orange curve). The measured capture time constant (gray dashed line) agrees consistently with the simulation and calculation at the measured Gd concentration (gray vertical dashed line) in each plot. The bands associated with the lines correspond to the respective uncertainties.}
    \label{fig:calc_g4_capt}
\end{figure}

Figure~\ref{fig:calc_g4_capt} shows that the capture time constant as a function of the Gd concentration from the output default Geant4 (red circle) and Geant4 with the thermal boost modification (blue box) in the SK-VI (left) and SK-VII periods, respectively.
Both the default and modified Geant4 give identical results to the analytical calculation (orange line) within 0.1\% over the displayed range of the Gd concentration.
The data estimations (gray lines)~\cite{cite:first, cite:second} correspond to the measurement of the Gd concentration in the water sampled from the SK detector (vertical) and the capture time constant measured in the SK data (horizontal), respectively.
It can be seen that the simulation result and the analytical calculation excellently reproduce the measured data in terms of the capture time constant. 

It is natural that the modification in \verb|G4ParticleHPThermalBoost.hh| should have a negligible effect on the capture time constant because Geant4 determines a type of neutron interaction in \verb|G4HadronicInteraction|, a base class of the \verb|G4ParticleHP| interaction classes, based on the cross section at a kinetic energy in the laboratory frame.
The capture time is defined when the \verb|G4ParticleHPCapture| is called from \verb|G4HadronicInteraction|, and then a type of nucleus capturing neutron is selected based on the cross section value with the thermal boost correction.
Therefore, the thermal boost modification should be visible in the H capture fraction instead of the capture time constant.

\subsection{H capture fraction}
\begin{figure}[h]
    \centering
    \includegraphics[width=0.49\linewidth]{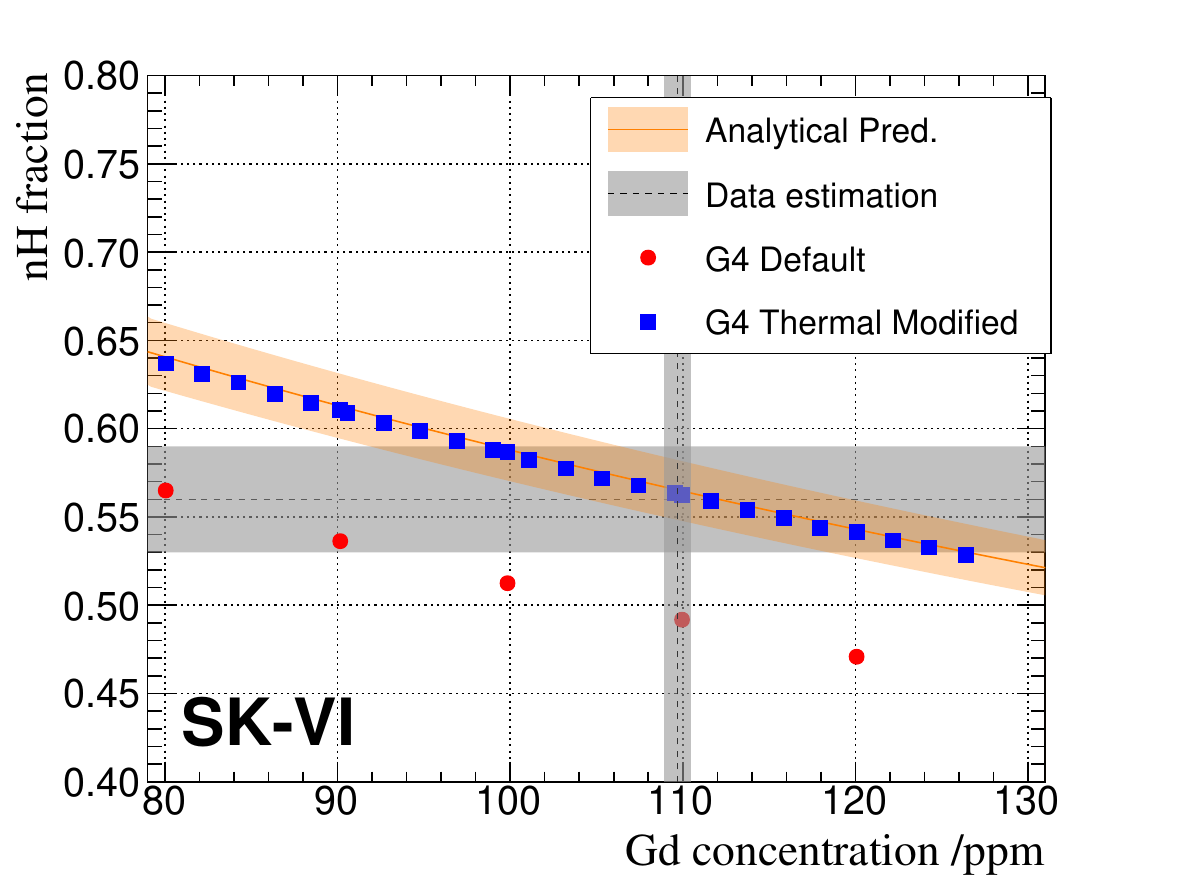}
    \includegraphics[width=0.49\linewidth]{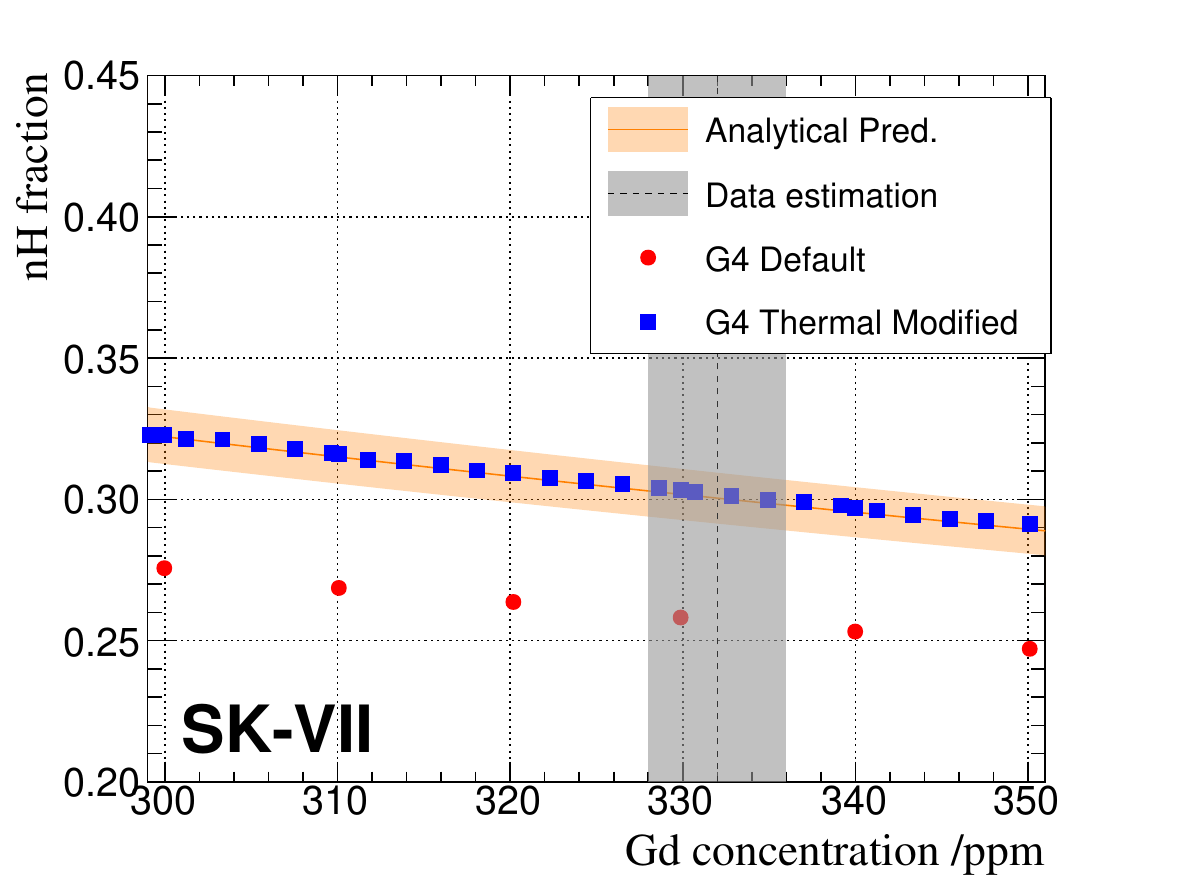}
    \caption{Hydrogen capture fraction as a function of Gd concentration around SK-VI 0.01\% (left) and SK-VII 0.03\% (right). The modified Geant4 (blue square) gives identical results to the analytical calculation (orange curve). On the other hand, the default Geant4 predicts a much lower H capture fraction in both plots (red circle). The measured H capture fraction (gray horizontal dashed line) agrees consistently with the modified Geant4 result and the calculation at the measured Gd concentration in the SK-VI phase. The bands associated with the lines correspond to the respective uncertainties.}
    \label{fig:calc_g4_frac}
\end{figure}

Figure~\ref{fig:calc_g4_frac} shows the H capture fraction as a function of the Gd concentration in a manner analogous to that of Fig.~\ref{fig:calc_g4_capt}. 
Obviously, the default Geant4 simulation (red circle) shows a +8\% deficit in the H capture fraction from the analytical calculation over the displayed Gd concentration in both plots.
In contrast, the simulation with the modified thermal boost (blue square) gives an identical result to the analytical calculation within less than 1\%.
The left plot, the SK-VI phase with Gd 0.01\%, shows that both the modified Geant4 simulation and the calculation agree consistently with the data estimated fraction $56 \pm 3$\%~\cite{cite:han_thesis} at $109.8 \pm 0.6$ ppm~\cite{cite:first}.
Therefore, it can be concluded that the thermal boost modification works appropriately and recovers the capability of predicting a neutron capture efficiency in Gd-loaded water.

Based on the modified simulation, the expected value of the H capture fraction in the SK-VII phase is $\sim$30\% at $332 \pm 4$ ppm~\cite{cite:second}.
Due to the higher Gd capture fraction in SK-VII, an underestimation of the H capture fraction should have less impact on the detection efficiency estimate.
However, the simulation with the modified thermal boost promises an evaluation of the neutron detection efficiency in the SK-Gd phases with a smaller systematic uncertainty on the estimation than ever before.
These expectations should be confirmed with the SK-Gd data in the near future.

\section{Discussion}
\label{sec:disc}
In contrast to the SK-Gd report, neutrino experiments using a Gd-loaded liquid scintillator (Gd-LS) have reported no evidence of the hydrogen capture fraction estimation malfunction in Geant4.
The Double Chooz experiment, a reactor neutrino oscillation experiment using 0.1w\% Gd-LS in France, shows a good agreement in the energy spectrum of the total neutron capture events (carbon as well as H and Gd) between the data and the MC simulation based on Geant4~\cite{cite:dchooz}.
The estimated Gd capture fractions are consistent within 2\% between the data and their MC, and it is less significant in their analysis because of the high Gd concentration~\cite{cite:konno_thesis}.
Accordingly, the Daya Bay experiment reported that the Gd capture fraction of their 0.1w\% Gd-LS detector predicted by Geant4.9 agreed with the measured value within 0.7\% accuracy~\cite{cite:dayabay}.
The short baseline neutrino oscillation experiment at J-PARC, JSNS${}^{2}$, uses the Gd-LS (0.1w\%) detector and reports that reproducibility of the neutron capture event spectrum in the ${}^{252}$Cf calibration data by their Geant4.10 based MC simulation~\cite{cite:hino_thesis}.
The experiments with a high concentration Gd-loaded detector succeeded in predicting neutron capture independent of the major release version of Geant4, as shown above.

The estimation of the neutron detection efficiency in the early SK-Gd phases suffered from the Geant4 implementation issue.
It has a greater impact when using the diluted Gd concentration, e.g., Gd 0.01\%, which gives approximately 50\% of the H capture fraction.
However, it would be difficult to find the issue if the Gd concentration of SK-Gd started at 0.1\%, as in the other neutrino experiments using high concentration Gd-LS or Gd-loaded water.

\subsection{Comments on the pure water phase}
\begin{figure}[h]
    \centering
    \includegraphics[width=0.7\linewidth]{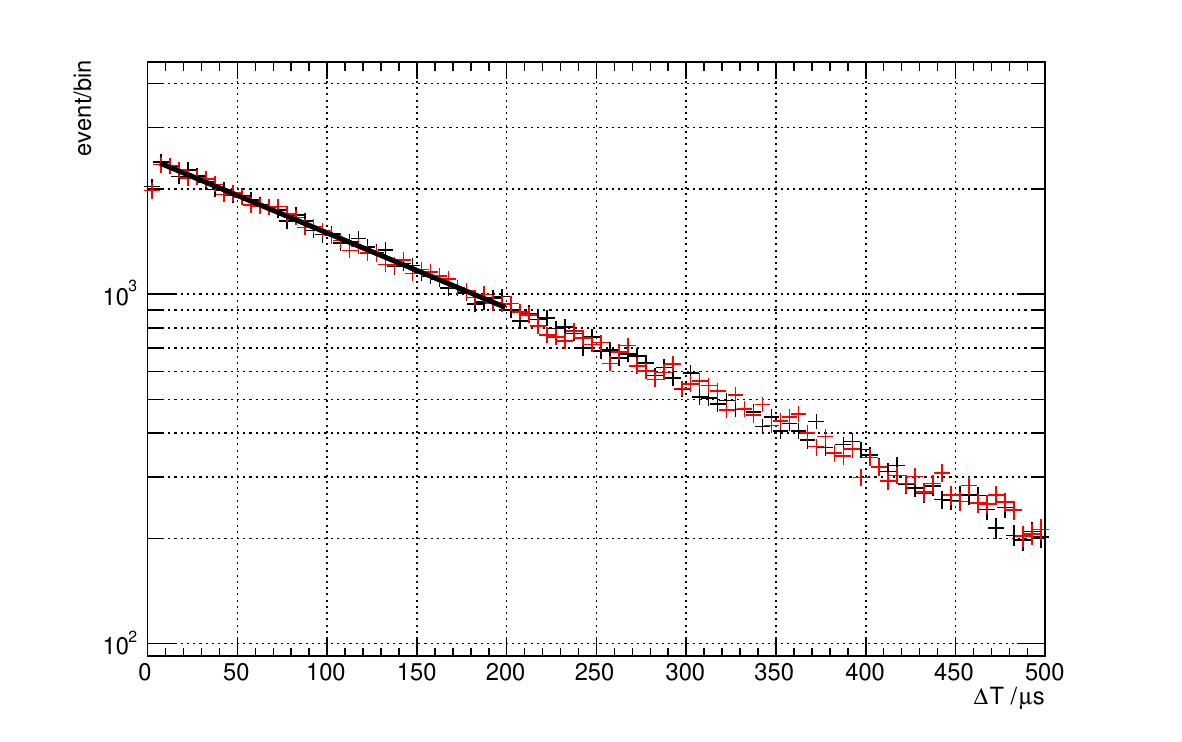}
    \caption{Neutron capture time distributions in pure water with the default Geant4 (black) and the modified Geant4 (red). As discussed in the main text, there is no change in the case of pure water because there are no competing capture nuclei other than hydrogen.}
    \label{fig:capt_purew}
\end{figure}

As discussed in Section~\ref{sec:result}, a neutron detection efficiency using neutron capture is greatly affected by the \verb|G4ParthicleHPThermalBoost| when multiple capture target nuclei are present in a detector material.
Thus, in cases where only one nucleus captures neutrons, such as the pure water phase in SK, the Geant4-based simulation adequately predicts the neutron capture constant and the detection efficiency. 
Figure~\ref{fig:capt_purew} shows the capture time distributions for the default Geant4 (black) and Geant4 with the modified thermal boost (red) in pure water, respectively.
As expected, they are in good agreement with each other.
We can conclude that the simulation result with the default Geant4 is valid in the pure water phase.

\section{Conclusion}
\label{sec:conc}
There have been reports showing a discrepancy in the neutron detection efficiency estimate between the data and the MC simulation in the early phases of SK-Gd, especially in the SK-VI 0.01\% Gd concentration.
The investigation of the neutron capture implementation in Geant4 showed that the thermal motion of a target capture nucleus considered in the thermal neutron capture was overestimated in the case of hydrogen.
As a result of the modification of the thermal motion calculation for hydrogen, the observables, e.g., capture time constant, and H capture fraction, were predicted to be identical to the analytical calculation at a given Gd concentration.
The Geant4 prediction with the modified thermal boost gives consistent values for both the capture time constant and the H capture fraction with the data estimates in SK-Gd. 
This modification is expected to reduce the systematic uncertainty associated with the neutron detection efficiency and to provide a safe prediction in analyses using neutron tags, such as the DSNB search in SK-Gd and experiments using low concentration metal-loaded liquid detectors.

\section*{Acknowledgments}
We are grateful to the Super-Kamiokande Collaboration for allowing us to perform the simulation study based on SKG4.
This work was supported by Japan MEXT KAKENHI Grant Number 24H02242 and 20K03986.




\end{document}